\documentclass[twocolumn, pra, showpacs]{revtex4}
\usepackage{graphicx}
\usepackage{dcolumn}
\usepackage{bm}
\usepackage{amsmath}

\begin{document}

\title{Ground-state properties of interacting two-component Bose gases in a
one-dimensional harmonic trap}
\author{Yajiang Hao and Shu Chen}
\email{schen@aphy.iphy.ac.cn}
\affiliation{Beijing National
Laboratory for Condensed Matter Physics, Institute of Physics,
Chinese Academy of Sciences, Beijing 100080, P. R. China}

\begin{abstract}
We study ground-state properties of interacting two-component
boson gases in a one-dimensional harmonic trap by using the exact
numerical diagonalization method. Based on numerical solutions of
many-body Hamiltonians, we calculate the ground-state density
distributions in the whole interaction regime for different atomic
number ratio, intra- and inter-atomic interactions.  For the case
with equal intra- and inter-atomic interactions, our results
clearly display the evolution of density distributions from a Bose
condensate distribution to a Fermi-like distribution with the
increase of the repulsive interaction. Particularly, we compare
our result in the strong interaction regime to the exact result in
the infinitely repulsive limit which can be obtained by a
generalized Bose-Fermi mapping. We also discuss the general case
with different intra- and inter-atomic interactions and show the
rich configurations of the density profiles.

\end{abstract}

\date{\today}
% 03.75.Mn Multicomponent condensates; spinor condensates
% 03.75.Hh Static properties of condensates,thermodynamical, statistical and structural properties.
%67.60.Bc Boson mixtures
%67.40.Db Quantum statistical theory; groundstate, elementary excitations
%64.75.+g Solubility, segregation, and mixing; phase separation
%\keywords{Bose mixture, Tonks Girardeau gases, phase separation}

\maketitle

\section{Introduction}

The realization of two-component Bose-Einstein condensates (BECs)
of trapped alkali atomic clouds \cite{Myatt,Erhard} has stimulated
the great interest in the multi-component atomic (or spinor)
gases. The competition between inter-species and intra-species
interactions gives rise to a richer structure of spinor
condensates than their single component counterparts. Many
theoretical investigations have focused on its stability, phase
separation, collective excitation, dynamics and other macroscopic
quantum phenomena \cite{Ho96,Ao,Pu,Cazalilla}. Moreover resonant
control of two-body interactions via Feshbach resonances in
ultracold gases makes it possible to study a two-component
condensate with a tunable interspecies interaction. On the other
hand, the experimental progress in the creation and manipulation
of ultracold atomic gases in effective one-dimensional (1D)
waveguides \cite {Stoeferle,Paredes,Toshiya} offers the
opportunity of studying the many-body physics of the 1D correlated
quantum gases even in the strong interaction limit. The
experimental realization of Tonks-Girardeau (TG) gases
\cite{Girardeau} in the strongly interacting limit of Bose gases
has been reported in two recent experiments
\cite{Paredes,Toshiya}.

For the 1D quantum gases \cite{Olshanii}, quantum fluctuations are
greatly enhanced and the effects of strong correlations become
very important \cite {Olshanii2,Petrov,Dunjko,Chen}. Although
two-component quantum gases have been the subject of intensive
theoretical interest \cite{Ho96,Ao,Pu,Cazalilla}, most of the
studies were carried out within the scheme of the mean field
theory which are proven insufficient for strong interactions
between particles. For the system in the strong interaction
regime, non-perturbation theory has to be exerted. For the 1D
homogeneous system with $\delta$-function interaction, the
two-component Bose gas with spin-independent s-wave scattering
(equal interspecies and intra-species interacting strength) is
known to be exactly solvable by the Bethe-ansatz method in the
full physical regime \cite{LiYQ}. The exact results have unveiled
that ferromagnetic order emerges in spinor Bose gases with
spin-independent interaction \cite{Eisenberg,LiYQ,Guan07,Fuchs}.
But for the atomic gases with arbitrary interacting strength (with
the SU(2) symmetry broken) or in an inhomogeneous trap, the system
is no longer integrable.

Despite no exact solution for the many-body systems with arbitrary
interacting strength in a harmonic trap available, it is still
possible to resort to numerical exact diagonalization method to
obtain the exact solution for small size system. Most recently,
some sophistical theoretical methods, such as the exact
diagonalization method \cite{Deuretzbacher,Yin} and multi-orbital
self-consistent Hartree method
\cite{Zoellner,Zoellner06,Cederbaum,Cederbaum06} have been applied
to study the single-component few-boson systems in a harmonic
trap. Due to properly accounting the many-body effects, such
methods work well even in strongly interacting regime and yield
consistent results with the exact results obtained by the
Bethe-ansatz method \cite{Hao06,Hao07}. In this work, we shall
study two-component boson systems with repulsive interactions and
focus on few-atom systems which can be treated exactly by exact
diagonalization method. One of the purpose of the present work is
to address the role of the correlation in the trapped interacting
two-component system under a well controllable way, in particular
in the strongly interacting limit where the approximate mean-field
methods are generally not able to give correct results. Our study
can display the transition from a weakly interacting condensate to
the strongly interacting limit in which the strongly repulsive
interactions between atoms plays a role of effective Pauli
principle. In this limit, we compare our result with the exact
result in TG-gas limit via a generalized Bose-Fermi mapping which
has been applied to the Bose-Fermi mixtures by Girardeau et. al
very recently \cite {Girardeau07}. We also address the case with
different inter-species and intra-species interactions.

Our paper is organized as follows. Section II introduces the model and gives
a brief introduction to the computational method. The subsequent section is
devoted to our results for the ground-state density distribution. A summary
is given in the last section.

\section{The microscopic model and the method}

Let us consider the two components of Bose gases confined in a
strongly anisotropic harmonic trap, where the transverse
confinement frequency $\omega _{\perp }$ is strong enough so that
the dynamics of Bose gases are effectively described by a 1D model
due to the radial degrees of freedom are frozen. We start with the
microscopic Hamiltonian of two species of Bosons
\begin{eqnarray}
H &=&\int dx\sum_{\alpha =1,2}\left\{ \Psi _\alpha ^{\dagger }(x)\left[ -%
\frac{\hbar ^2}{2m_\alpha }\frac{\partial ^2}{\partial x^2}+V_\alpha
(x)\right] \Psi _\alpha (x)\right.  \nonumber \\
&&\left. +\frac{g_\alpha }2\Psi _\alpha ^{\dagger }(x)\Psi _\alpha ^{\dagger
}(x)\Psi _\alpha (x)\Psi _\alpha (x)\right\}  \nonumber \\
&&+g_{12}\int dx\Psi _1^{\dagger }(x)\Psi _2^{\dagger }(x)\Psi _2(x)\Psi
_1(x)
\end{eqnarray}
with $V_\alpha (x)=\frac 12m_\alpha \omega _\alpha ^2x^2$ ($\alpha =1,2$)
being the trap potential felt by the $\alpha $-component bosons with mass $%
m_\alpha $ and longitudinal trap frequency $\omega _\alpha $. Here
$\Psi _\alpha (x)$ denotes the field operator of the $\alpha
$-component bosons for the longitudinal degree of freedom. In the
following we will focus on the two-component Bose gases composed
of two hyperfine state of the same atoms such that $m_1=m_2=m$ and
it is assumed that both of them feel the same trap frequency,
i.e., $\omega _1=\omega _2=\omega $. The effective intra- and
inter-atomic interaction strengths are denoted as $g_\alpha
=-2\hbar ^2/(ma_{1d}^\alpha )$ ($\alpha =1,2$) and $g_{12}=-2\hbar
^2/(ma_{1d}^{12})$, respectively. They are related to the three dimensional $%
s$-wave scattering length between two atoms ($a_s^1$ and $a_s^2$
being the intra-atomic scattering length and $a_s^{12}$ being the
inter-atomic scattering length) by \cite{Olshanii,Olshanii2}
\[
a_{1d}^\alpha =-l_{\perp }\left( \frac{l_{\perp }}{a_s^\alpha }-\frac{\left|
\zeta (1/2)\right| }{\sqrt{2}}\right) ,
\begin{array}{ll}
&
\end{array}
\alpha =1,2,12
\]
with $l_{\perp }=\sqrt{\hbar /m\omega _{\perp }}$ the transverse
characteristic oscillator length.

Expanding the operator $\Psi _\alpha (x)$ in the basis of harmonic
oscillator functions (orbital) $\phi _i(x)$ with $i$ the orbital index
\[
\Psi _\alpha (x)=\sum_{i=0}^\infty \phi _i(x)\hat{b}_{i\alpha },
\]
we get the many-body Hamiltonian in terms of the creation and destruction
operators $\hat{b}_{i\alpha }^{\dagger }$ and $\hat{b}_{i\alpha }$ of the
axial harmonic oscillator
\begin{eqnarray}
H &=&\sum_{\alpha =1,2}\left[ \sum_i\mu _i\hat{b}_{i\alpha }^{\dagger }\hat{b%
}_{i\alpha }+\frac{U_\alpha }2\sum_{i,j,k,l}I_{i,j,k,l}\hat{b}_{i\alpha
}^{\dagger }\hat{b}_{j\alpha }^{\dagger }\hat{b}_{k\alpha }\hat{b}_{l\alpha
}\right]  \nonumber \\
&&+U_{12}\sum_{i,j,k,l}I_{i,j,k,l}\hat{b}_{i1}^{\dagger }\hat{b}%
_{j2}^{\dagger }\hat{b}_{k2}\hat{b}_{l1}
\end{eqnarray}
with $\mu _i=i+\frac 12$. Here we have rescaled the energy and
length in units of $\hbar \omega $ and $l_0=\sqrt{\hbar /m\omega
}$. The parameters $U_\alpha $ ($\alpha =1,2,12$) are
dimensionless interaction constants and $
I_{ijkl}=l_0\int_{-\infty }^\infty dx\phi _i(x)\phi _j(x)\phi
_k(x)\phi _l(x) $ are dimensionless integrals. The ground state of
the Hamiltonian can be obtained by diagonalizing it in the Hilbert
space spanned by the eigenstates of the harmonic oscillator.
Although the exact diagonalization methods can in principle solve
the ground-state problem of the many-body Hamiltonian exactly, one
has to truncate the set of basis functions to the lowest L
orbitals for reasons of numerical feasibility. The dimension of
the space should be $C_{N_1+L-1}^{N_1}\times C_{N_2+L-1}^{N_2}$ if
$N_1$ 1-component atoms and $N_2$ 2-component atoms are populated
on $L$ orbitals. In principle the exact ground state of the
many-body Hamiltonian can be obtained even in the strongly
interacting regime as long as $L$ is large enough. Our goal is to
investigate the ground states of the interacting system for all
relevant interaction strengths in a numerically exact way which
however restricts our study to small particle system. We note that
either the effort to extend the exact diagonalization method to
solve the larger system or to get a well controllable numerical
result in the strongly interacting limit is a highly challenging
and time-consuming task as what have been displayed in the works
for the single-component model
\cite{Deuretzbacher,Zoellner,Zoellner06}.

\section{Ground state properties}

In the following we will investigate both the density distribution of the
many-body ground state of each component and the total distribution, which
is given by
\[
\rho (x)=\sum_{\alpha =1}^2\rho _\alpha (x)
\]
with
\begin{eqnarray}
\rho _\alpha (x) &=&\left\langle \text{GS}\right| {\Psi }_\alpha ^{\dagger
}(x){\ \Psi }_\alpha (x)\left| \text{GS}\right\rangle  \nonumber \\
&=&\sum_{i,j}\phi _i^{*}(x)\phi _j(x)\left\langle \text{GS}\right| \hat{b}%
_{i\alpha }^{\dagger }\hat{b}_{j\alpha }\left| \text{GS}\right\rangle .
\end{eqnarray}
The ground state $\left| \text{GS}\right\rangle $ can be obtained
by numerically diagonalizing the Hamiltonian in the reduced
Hilbert space which is typically composed of $10^5$ basis states
in our calculation. To guarantee the numerical exactness, in the
following we will focus on the system of four bosons with the
maximum energy cutoff of $30 \hbar \omega$ corresponding to the
number of orbitals L=30.
\begin{figure}[tbp]
\includegraphics[width=3.5in]{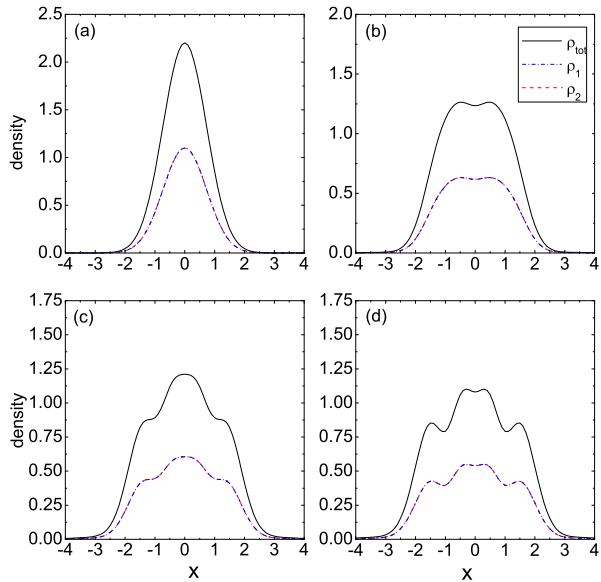}\newline
\caption{Density distribution of the ground state for $N_1=N_2=2$. (a) $%
U_1=U_2=U_{12}=0.1$; (b) $U_1=U_2=U_{12}=3.0$; (c)
$U_1=U_2=U_{12}=7.0$; (d) $U_1=U_2=U_{12}=10.0$.} \label{fig1}
\end{figure}

We first consider the case with the intra- and inter-atomic
interactions being same, i.e., $U_1=U_2=U_{12}$. In Fig. 1, we
display the density distributions of two components of Bose gas in
a harmonic trap when both components have same particle numbers.
As shown in the figures each component of the Bose gas has the
same distribution in the whole regime of interaction. In the weak
interaction regime, the Bose gas exhibits a Gauss-like
distribution, which implies that the many-boson system can be
approximately described by a single wave function, i.e., all the
bosons condense into the same single state. In the scheme of the
mean-field theory, the single wave function is obtained by solving
Gross-Pitaevskii equations. In the limit of $U_\alpha =0$ ($\alpha
=1,2,12$) all atoms occupy the lowest orbital and the ground state
takes the form of
\[
\psi (x_1^1,\ldots ,x_{N_1}^1;x_1^2,\ldots ,x_{N_2}^2)=\prod_{i=1}^{N_1}\phi
_0(x_i^1)\prod_{j=1}^{N_2}\phi _0(x_j^2),
\]
which can be expressed as $|\text{GS}\rangle =(b_{01}^{\dagger
})^{N_1}(b_{02}^{\dagger })^{N_2}|0\rangle $ in our Hilbert space.
With the increase of the atomic interaction, the density
distribution becomes broader and flatter. When it reaches the
strongly interacting regime, four peaks appear at $U_\alpha =10$
as shown in the Fig.1d. The emergence of density oscillations is a
signal of ``fermionization", i.e., the density profile shows
fermion-like distribution \cite{Girardeau07,Girardeau}. In this
case, two bosons would avoid to occupying the same orbital and
thus the bosons occupy the orbitals one by one similar to the
fermions. Therefore the mean-field theory based on the single wave
function approximation is not a suitable theory in the strongly
interacting limit. To capture the character of density
oscillations, more sophistical multi-orbital mean-field theory is
required \cite{Zoellner,Zoellner06,Cederbaum,Cederbaum06}.
\begin{figure}[tbp]
\includegraphics[width=3.5in]{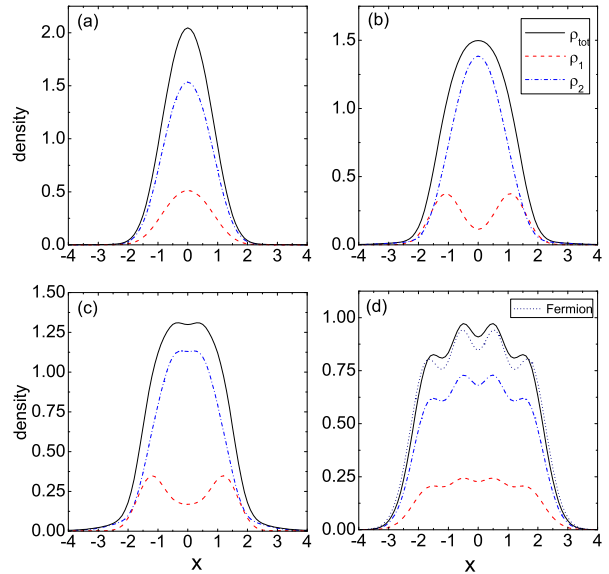}\newline
\caption{Density distribution of the ground state for $N_1=1$ and $N_2=3$.
(a) $U_1=U_2=U_{12}=1.0$; (b) $U_1=U_2=U_{12}=3.0$; (c) $U_1=U_2=U_{12}=5.0$%
; (d) $U_1=U_2=U_{12}=10.0$.}
\label{fig2}
\end{figure}

For the imbalanced case with $N_1\neq N_2$, the total density distribution
displays the similar crossover behavior as the case with $N_1=N_2$. As shown
in Fig.2, the halfwidth of the total density distributions becomes larger
and larger with the increase of interaction strength, followed by the
emergence of an obvious spatial oscillation structure at $U_\alpha =10$.
What special here is the density distribution for each component is no
longer same in the whole interacting regime. In the weakly interacting
regime, each component has similar density profile. While in the middle
regime, the component of the Bose mixture behaves in different way. At $%
U_\alpha =3$, two obvious peaks are observed for the component
with less atoms whereas the total density shows no oscillation.
With further increasing the interaction strength to $U_\alpha =5$,
the total density distribution begins to oscillate with two peaks
emerging. When reaching the strongly interacting regime with
$U_\alpha =10$, four obvious peaks are observed for both
components of the Bose mixture. In this limit each component has
similar density profile again: the density distribution of the
component with $N_\alpha$ atoms is ${N_\alpha}/{(N_1+N_2)}$ of the
total density distribution with $N_1+N_2$ peaks. As we shall
discuss latter, this is consistent with the result in the limit of
infinite interaction, where both the total distribution of two
species and the respective distribution of them behave like
noninteracting spinless fermions.
\begin{figure}[tbp]
\includegraphics[width=3.5in]{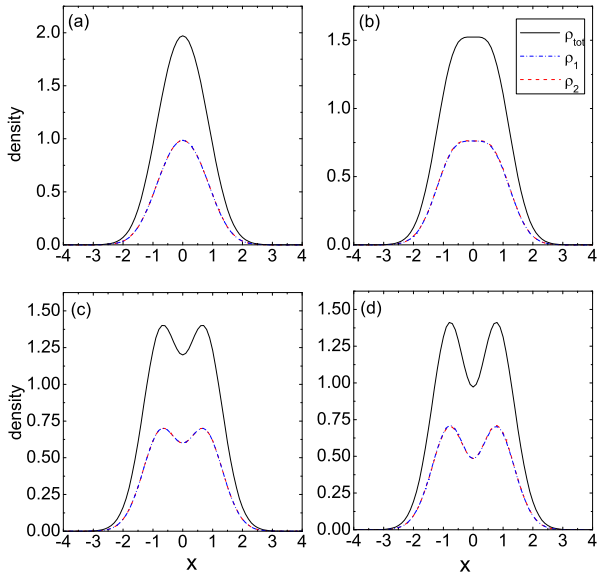}\newline
\caption{Density distribution of the ground state for $N_1=N_2=2$
and $ U_1=U_2=0.1$. (a) $U_{12}=0.1$; (b) $U_{12}=3.0$; (c)
$U_{12}=5.0$; (d) $ U_{12}=10.0$.} \label{fig3}
\end{figure}

In the limit of $U_\alpha =\infty $ , we can apply the generalized
Bose-Fermi mapping method \cite{Girardeau07,Deuretzbacher08} to
construct the ground state wavefunction of the Bose mixture.
According to Girardeau's argument \cite {Girardeau}, in the
infinite interacting limit the interactions can be replaced by
constraints that the many-body wave function vanishes at all
collision points \cite {Girardeau07,Girardeau,Girardeau02}. It
follows that the Girardeau-type ground state wavefunction has the
following form
\begin{eqnarray}
&&\psi (x_1^1,\ldots ,x_{N_1}^1;x_1^2,\ldots ,x_{N_2}^2)  \nonumber \\
&=&A(x_1^1,\ldots ,x_{N_1}^1,x_1^2,\ldots ,x_{N_2}^2)\times   \nonumber \\
&&\psi _F(x_1^1,\ldots ,x_{N_1}^1;x_1^2,\ldots ,x_{N_2}^2)
\end{eqnarray}
with the antisymmetric function $A(x_1,x_2,\ldots
,x_N)=\prod_{1\leq i,j\leq N}\text{sgn}(x_i-x_j)$ and $\psi
_F(x_1^1,\ldots ,x_{N_1}^1;x_1^2,\ldots ,x_{N_2}^2)$ is a Slater
determinant of $N=N_1+N_2$ orthonormal orbitals $ \phi
_0(x),\cdots ,\phi _{N-1}(x)$ occupied by both components of
bosons. Explicitly, we have
\[
\psi _F(x_1,\ldots ,x_N)=(N!)^{-\frac 12}det[\phi
_j(x_i)]_{i=1,2,\ldots ,N}^{j=0,1,\ldots ,N-1}
\]
with $x_1,\ldots ,x_N$ denoting $x_1^1,\ldots
,x_{N_1}^1;x_1^2,\ldots ,x_{N_2}^2$. Now it is straightforward to
get
\begin{equation}
\rho _\alpha (x)=\frac{N_\alpha }N\rho _F(x)
\end{equation}
with $\rho _F(x)=\sum_{i=0}^{N-1}\left| \phi _i(x)\right| ^2,$
i.e., the densities of both components are proportional to the
density of a harmonically trapped TG gas of $N$ bosons which is
identical to the density of $N$ free fermions. In Fig. 2d, we
compare our numerical result with the exact result in the TG
limit. It is clear that the numerical result complies with the
exact results very well which implies that the system has entered
the strongly interacting regime at this parameter point. Our
numerical results provide a clear evidence that the state obtained
by Girardeau-type construction \cite{Girardeau07} remains to be a
good approximation for large but finite repulsion.
\begin{figure}[tbp]
\includegraphics[width=3.5in]{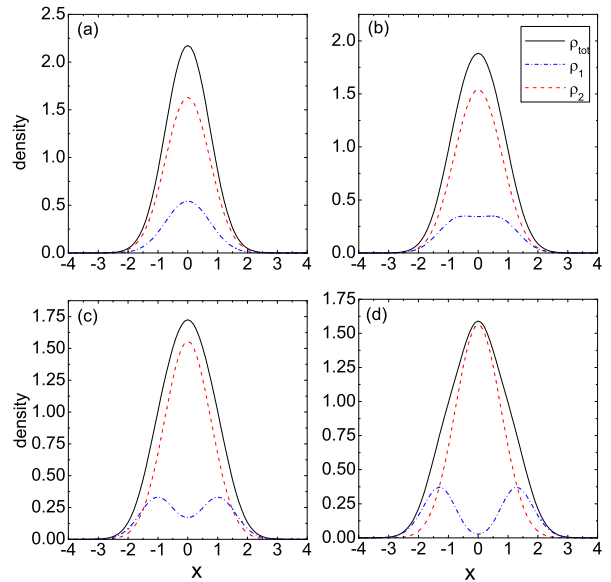}\newline
\caption{Density distribution of the ground state for $N_1=1$, $N_2=3$ and $%
U_1=U_2=0.1$. (a) $U_{12}=0.1$; (b) $U_{12}=1.0$; (c)
$U_{12}=2.0$; (d) $U_{12}=10.0$.} \label{fig4}
\end{figure}

Now we discuss the general case with different intra- and
inter-atomic interactions. First we calculate the density
distributions for the equal-mixing Bose mixture with equal weak
intra-component repulsions. For convenience, we take $U_1=U_2=0.1$
and $U_{12}\neq U_1,U_2$. Density distributions for different
inter-component interactions are shown in Fig. 3.  In all the
interaction regime, we have $\rho _1 (x)=\rho _2 (x) =\frac{1 }2
\rho(x)$. With the increase of $U_{12}$, the density distribution
becomes broader. For $U_{12}=5$, two peaks occur. With further
increase of inter-component interaction, only two peaks in the
density distribution are observed even for $U_{12}=10$. Due to the
intra-atomic interaction is weak, each component of atoms may form
a condensate. Similar results are found for the system with weak
intra-component repulsions even if $ U_1\neq U_2$. The difference
from the plot of the case of $U_1=U_2$ is that the component with
stronger intra-atomic interaction distributes in broader regime
than that with weaker interaction.

For the case of $N_1\neq N_2$ with weak intra-atomic interaction $
U_1=U_2=0.1$, density distributions for different inter-component
interactions are displayed in the Fig. 4. Similar to the
equal-mixing case, the total density distribution becomes flatter
with the increase of $U_{12}$. Although two obvious peaks occur
for the component with less atoms when $U_{12}=2.0$,  no obvious
peak is detected in the total density profiles even for
$U_{12}=10.0$. When the inter-atomic interaction is strong, the
density for the different component has different spacial
distribution and gets demixed in order to lower the inter-species
interaction energy. As shown in the figures, the component with
more atoms stays in the middle of the trap and the component with
less atoms stays in the regime away from the middle of the trap.

Finally, we study the change of density distributions versus
different inter-component interactions for the Bose mixture with
strong intra-atomic interactions. For convenience, we consider the
imbalanced case with $N_1=3$ and $N_2=1$ and we take equal strong
intra-component repulsions $U_1=U_2=10.0$. For the vanishing
inter-component interactions, the atoms of two species already lie
in the strongly interacting regime and the density distribution
for each component gets the Fermi-like feature. The distribution
of $\rho _1 (x)$ displays three peaks and the distribution of
$\rho _2 (x)$ displays one peak. As the inter-atomic interaction
increases, the distribution of each component changes continuously
as shown in Fig. 5. When $U_{12}=3.0$, the central peak of the
first component is suppressed and the density profiles get
broader. At $U_{12}=8.0$, four peaks occur for the first component
whereas the density for the second component get more flatter
accompanying with two peaks discernable. With further increase to
$U_{12}=10.0$, both the total density distribution and the
distribution of each component show $N_1 + N_2$ peaks and the
latter takes on the same distribution normalizing to their own
particle numbers $N_{\alpha}$.
\begin{figure}[tbp]
\includegraphics[width=3.5in]{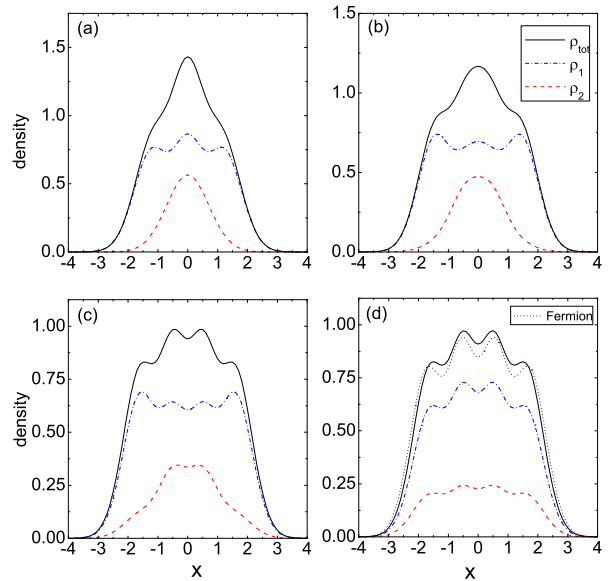}\newline
\caption{Density distribution of the ground state for $N_1=3$,
$N_2=1$ and $ U_1=U_2=10.0$. (a) $U_{12}=0.0$; (b) $U_{12}=3.0$; (c)
$U_{12}=8.0$; (d) $ U_{12}=10.0$.} \label{fig5}
\end{figure}

\section{Summary}
In summary, we have investigated the ground-state properties of
two components of Bose gases composed of two hyperfine states of
the atoms. With the numerical diagonalization method the density
distributions are obtained, which display obviously different
properties for different parameters in the system. For the system
with equal intra-component and inter-component interactions, the
total density distribution shows similar properties to the
single-component Bose gases with $N_1+N_2$ atoms. The evolution
from weakly interacting condensate to strongly interacting Tonks
gas is shown with the increasing atomic interaction. In the weakly
interacting regime the Bose mixture displays Gauss-like
distribution whereas in the strongly interacting regime the
density distribution gets Fermi-like features. Our numerical
result is consistent with the exact result in the infinite limit
and thus provides numerical evidence that the Bose mixture can be
approximately described by the TG gas even for the system with
strong but finite repulsion. We also discuss the general case with
different intra-component and inter-component interactions and
display the evolution of density distributions from weak to strong
inter-component interaction regime for both cases with weak and
strong intra-component interactions.  Despite the size limitation
of the exact diagonalization method used in the present work, our
results are no doubt very meaningful because it provides us
important insights into the quantum many-body physics beyond
various approximate effective-field approaches.

{\it Note added:} After finishing the present work \cite{Hao08},
we noticed a related work by Z\"{o}llner {\it et. al}
\cite{Zoellner08}, in which similar issue was discussed by the
multi-orbital self-consistent Hartree method.

\begin{acknowledgments}
S.C. is supported by NSF of China under Grant No. 10574150, MOST
grant 2006CB921300 and programs of Chinese Academy of Sciences.
\end{acknowledgments}

\end{document}